# Robust Estimation of Reactive Power for an Active Distribution System

Zhengshuo Li, Jianhui Wang, Hongbin Sun, Feng Qiu, Qinglai Guo

*Abstract*—Increasing distributed energy resources (DERs) may result in reactive power imbalance in a transmission power system (TPS). An active distribution power system (DPS) having DERs reportedly can work as a reactive power prosumer to help balance the reactive power in the TPS. The reactive power potential (RPP) of a DPS, which is the range between the maximal inductive and capacitive reactive power the DPS can reliably provide, should be accurately estimated. However, an accurate estimation is difficult because of the network constraints, mixed discrete and continuous variables, and the nonnegligible uncertainty in the DPS. To solve this problem, this paper proposes a robust RPP estimation method based on two-stage robust optimization, where the uncertainty in DERs and the boundary-bus voltage is considered. In this two-stage robust model, the RPP is pre-estimated in the first stage and its robust feasibility for any possible instance of the uncertainty is checked via a tractable problem in the second stage. The column-and-constraint generation algorithm is adopted, which solves this model in finite iterations. Case studies show that this robust method excels in yielding a completely reliable RPP, and also that a DPS, even under the uncertainty, is still an effective reactive power prosumer for the TPS.

## I. INTRODUCTION

INCREASING distributed energy resources (DERs) may result in a shortage of reactive power source in transmission power systems (TPSs), e.g., in Germany, because "conventional power plants that provide reactive power for the TPS will partially be shut down" with the increase in DERs [1]. To balance regional reactive power and to avoid long-distance transmission of reactive power from other regions, transmission system operators (TSOs) need to find alternative regional reactive power sources.

In these circumstances, a distribution power system (DPS) is reportedly a potential reactive power prosumer to balance the reactive power in the upstream TPS [1]-[5]. This assertion is supported by the facts that 1) modern DPSs that are equipped with switchable capacitor bank shunts, online load tap changers (OLTCs), and even SVC/SVGs are typically able to regulate the reactive power at the coupling point, or *boundary bus*, of the DPS and the TPS, and moreover that 2) the DERs in the DPS are usually technically capable of injecting reactive power (*capacitive*) into and absorbing the excessive reactive power (*inductive*) from the grid, which further enhances the DPS's controllability over the boundary-bus reactive power. Thus, a DPS can intentionally provide inductive and/or capacitive reactive power to balance the reactive power in the TPS. Preliminary studies [1]-[5] have demonstrated the positive effect of using a DPS as a reactive power prosumer.

However, a remaining issue in the current research is how a distribution system operator (DSO) accurately estimates the DPS's *reactive power potential* (RPP), that is, "the range between the maximal inductive and capacitive reactive power at the boundary bus" as defined in [1]. The estimated RPP should be known by the TSO before he/she decides how much reactive power would be required from the DPS. For example, in [6], for the 10-minute real-time dispatch, a DSO should update the TSO on the newest RPP every ten minutes, and this RPP should be *reliable* in the sense that any value in the range of this RPP can actually be realized via the available controls on the DPS side. Otherwise, the TSO may require an amount of reactive power that cannot actually be delivered by the DPS, and consequently the issue of regional reactive power imbalance will arise, followed by other unwanted consequences.

It is fair to say that an accurate estimation of a DPS's RPP is technically difficult, because the operational constraints of the network, e.g., nodal voltage limits and power flow equations, and the constraints regarding the controllable devices, e.g., the OLTC, switchable shunts, and DERs, have to be considered. Moreover, for wind or solar DERs, the uncertainty in their active power, which also affects the RPP, should be considered. In addition, at the moment of computing the RPP, a DSO is usually uncertain about the boundary-bus voltage. That uncertainty occurs because this voltage setpoint, which is typically decided by the TSO, may change from the current value afterwards, and also because in practice the real-time voltage may fluctuate around, rather than be equal to, the setpoint. Hence, the RPP estimation problem involves uncertainty and discrete variables as well as complex operational constraints.

In the literature, a commonly used approach to estimating the maximal inductive and/or capacitive reactive power at the boundary bus is to solve one or two deterministic DPSs' optimal power flow (OPF) problems [3]-[5],[7],[8]. For tractability, linearized three-phase power flow is considered in [8] to compute the maximal capacitive reactive power, and the single-phase counterpart is adopted in [3],[4] to compute the RPP. In addition, reliance on the precondition that there are no dis-

This work is supported by the China Postdoctoral Science Foundation under Grants 2016M600091 and 2017T100078.

Zhengshuo Li is with Tsinghua-Berkeley Shenzhen Institute (TBSI), Tsinghua University, Shenzhen, Guangdong, 518055, China (email: shuozhengli@sina.com). Jianhui Wang is with Southern Methodist University, Dallas, TX, USA and Argonne National Laboratory, Argonne, IL, USA (email: jianhui.wang@ieee.org). Feng Qiu is with Argonne National Laboratory, Argonne, IL, USA. Hongbin Sun and Qinglai Guo are with Tsinghua University, Beijing, 100084, China (e-mail: shb@tsinghua.edu.cn).

crete control variables in the DPS [9] presents a method to draw the feasible region of the boundary-bus active and reactive power by solving a series of OPF problems. However, these deterministic methods (DMs) may not be suitable for computing the RPP when the uncertainty in DERs and boundary-bus voltage is considered. To handle the uncertainty, Monte Carlo simulation and OPF can be combined to evaluate the RPP [1], but this strategy is usually computationally expensive and thus might be inappropriate for online application. In [10], the authors evaluated the worst-case scenario maximal inductive and capacitive boundary-bus reactive power, but the worst-case scenario was found heuristically, which might decrease the accuracy of the result. Moreover, [10] also failed to consider the impact of the uncertainty in the boundary-bus voltage.

To resolve the above issues, this paper proposes a two-stage robust-optimization (RO)-based RPP estimation method, where the uncertainty in the DERs and the boundary-bus voltage is considered. Relative to the aforementioned methods, this robust method has the following advantages: First, the two-stage RO formulation closely fits the physical structure of the RPP estimation problem: a DSO first pre-estimates an RPP (defined as the first-stage problem, or FSP), then checks whether any reactive power in this range is realizable in the presence of uncertain DERs and boundary-bus voltage (defined as the second-stage problem, or SSP) and modifies the result of the FSP if necessary. Note that the worst-case scenario here is found via optimization techniques rather than heuristically. Second, one does not need a detailed probability function of the uncertainty variables, which is usually unachievable in practice. Third, in comparison with the Monte Carlo simulation and stochastic programming techniques, RO is typically more efficiently solvable and thus promising for power system applications, as has been demonstrated in studies on unit commitment (e.g., [11],[12]), economic dispatch (e.g., [13]-[15]), OPF (e.g. [16]-[18]), etc. Indeed, if the uncertainty set is a polyhedron, a column-and-constraint generation (C&CG) algorithm [19] yields the solution to a two-stage RO problem in finite iterations.

In short, relative to the other studies on RPP estimation, this paper has two distinct differences: 1) it considers the uncertainty in the DERs and the boundary-bus voltage[1], and coordinates the discrete and continuous variables to yield a maximal and *reliable* RPP; and 2) unlike heuristic worst-case scenario searching [10] or scenario-sampling techniques, this RO-based method accurately checks whether the RPP in the FSP is completely reliable. Our method also enjoys the three advantages summarized in the last paragraph.

The remainder of this paper will be arranged as follows: To facilitate understanding of this robust method, a deterministic DPS's operation model is first presented in Section II. Then, the robust model is described in Section III and a C&CG-based solution strategy is shown in Section IV. Next, case studies are presented in Section V and the conclusions are summarized in Section VI.

---

[1] Note that, also unlike other robust DPSs' OPF studies such as [17], the uncertainty in the boundary-bus voltage is considered here, which, though it makes the model more complex, improves the accuracy of the estimation.

## II. DETERMINISTIC DPS'S OPERATION MODEL

In this section, the famous DistFlow equations [20], which are a single-phase power flow model, are used to formulate the DPS's power flow equations, because under certain conditions DistFlow can be convexified into a second-order cone (SOC) model that facilitates solving a two-stage RO problem ([17],[21]). Although DistFlow is accurate only for a three-phase balanced DPS, it is still a popular tool in abundant studies on a DPS that can be deemed nearly three-phase balanced. Since this is typically the case for a high-voltage DPS that is directly interfaced with the TPS, we adopt the DistFlow model here. As will be seen in the sequel, however, our method is also applicable to a problem with linearized single- or three-phase power flow equations.

Suppose there are $N+1$ nodes and $N$ branches in a radial DPS and the root node, i.e. boundary bus, is marked with no. 1. The sets of the nodes and branches are denoted by $\mathcal{B}$ and $\mathcal{E}$, respectively. Moreover, $\mathcal{E} = \mathcal{T} \cup \mathcal{T}^c$ where $\mathcal{T}$ is the set of the branches with an OLTC (the model for which is shown in Fig. 1) and $\mathcal{T}^c$ the remainder. The DistFlow model is given below:

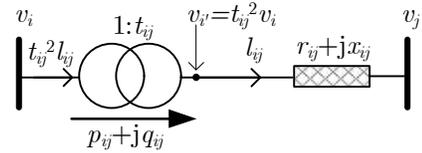

Fig. 1. The OLTC model used in this paper.

$$p_j = p_{G,j} - p_{D,j} = \sum_{k \in \mathcal{T}(j)} p_{jk} - \sum_{i \in \mathcal{H}(j)} \left( p_{ij} - r_{ij} l_{ij} \right) \quad \forall j \in \mathcal{B} \quad (1a)$$

$$q_j = q_{G,j} + q_{C,j} - q_{D,j} = \sum_{k \in \mathcal{T}(j)} q_{jk} - \sum_{i \in \mathcal{H}(j)} \left( q_{ij} - x_{ij} l_{ij} \right) \forall j \in \mathcal{B} \quad (1b)$$

$$\forall (i,j) \in \mathcal{T}^c : \begin{cases} v_j = v_i - 2\left( r_{ij} p_{ij} + x_{ij} q_{ij} \right) + \left( r_{ij}^2 + x_{ij}^2 \right) l_{ij} \\ p_{ij}^2 + q_{ij}^2 = v_i l_{ij} \text{ or } \left\| 2p_{ij}, 2q_{ij}, l_{ij} - v_i \right\|_2 = l_{ij} + v_i \end{cases} \quad (1c)$$

$$\forall (i,j) \in \mathcal{T} : \begin{cases} v_j = t_{ij}^2 v_i - 2\left( r_{ij} p_{ij} + x_{ij} q_{ij} \right) + \left( r_{ij}^2 + x_{ij}^2 \right) l_{ij} \\ p_{ij}^2 + q_{ij}^2 = t_{ij}^2 v_i l_{ij} \text{ or } \left\| 2p_{ij}, 2q_{ij}, l_{ij} - t_{ij}^2 v_i \right\|_2 = l_{ij} + t_{ij}^2 v_i \end{cases} \quad (1d)$$

$$v_1 = v^{\text{set}}, \text{ and for } \forall j \in \mathcal{B} \setminus \{1\}: \underline{v}_j \leq v_j \leq \overline{v}_j \quad (1e)$$

$$0 \leq l_{ij} \leq \left| \overline{I}_{ij} \right|^2 \quad \forall (i,j) \in \mathcal{E} \quad (1f)$$

where $p_j$, $q_j$ and $v_j$ denote the active and reactive power injection and the square of the voltage at node $j$, respectively; $p_{G,j}$ and $q_{G,j}$ (resp. $p_{D,j}$ and $q_{D,j}$) are the active and reactive power generation (resp. consumption) at node $j$, respectively; $q_{C,j}$ denote the reactive power from the capacitor shunt at node $j$; $p_{ij}$, $q_{ij}$ and $l_{ij}$ denote the active and reactive power and the square of the current flowing through branch $(i,j)$, respectively; $r_{ij}$ and $x_{ij}$ are the resistance and impedance of branch $(i,j)$; $t_{ij}$ is the tap ratio of the OLTC; $\mathcal{H}(j)$ and $\mathcal{T}(j)$ denote the sets of the father and child nodes of node $j$, respectively; and $v^{\text{set}}$ is the square of the voltage at the boundary bus, which is set by the TSO. The operational limits of the squares of the branch current and nodal voltages are $|I_{ij}|^2$, $\bar{v}_j$ and $\underline{v}_j$, respectively.

As for a DER at node $j$, its $p_{G,j}$ and $q_{G,j}$ typically should stay within the operational region $\sqrt{p_{G,j}^2 + q_{G,j}^2} \leq S_{G,j}$, as shown in Fig. 2, which can be further linearized as follows:

$$\begin{cases} p_{G,j} + q_{G,j} \leq \sqrt{2} S_{G,j},\ p_{G,j} - q_{G,j} \leq \sqrt{2} S_{G,j} \\ -S_{G,j} \leq q_{G,j} \leq S_{G,j},\ 0 \leq p_{G,j} \leq S_{G,j} \end{cases}. \quad (2)$$

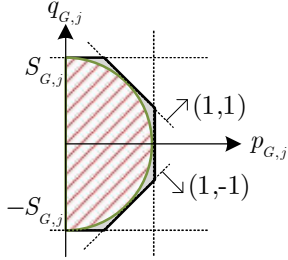

Fig. 2. The linearization of the operational constraints of a DER.

As for an SVC/SVG at node $j$, which is assumed to be continuously regulatable [17], $q_{G,j}$ should be constrained by

$$-S_{G,j} \leq q_{G,j} \leq S_{G,j}. \quad (3)$$

As for the capacitor shunt at node $j$, assume its capacitance $CP_j$ has $\mathrm{K}_{C,j} + 1$ possible values, $0, CP_{j,1}, \ldots, CP_{j,\mathrm{K}_{C,j}}$, via switching different capacitor banks. With $\mathrm{K}_{C,j}$ binary variables denoted by $b_{C,j,\kappa}$, $q_{C,j}$ can be formulated as

$$\begin{cases} q_{C,j} = \sum_\kappa CP_{j,\kappa} b_{C,j,\kappa} v_j \\ \sum_\kappa b_{C,j,\kappa} \leq 1,\ b_{C,j,\kappa} \in \{0,1\} \end{cases}. \quad (4a)$$

If both $v_j$ and $b_{C,j,\kappa}$ are optimal variables, then each bilinear term $b_{C,j,\kappa} v_j$ should be replaced by an auxiliary variable $z_{C,j,\kappa}$ subject to the McCormick's constraints, and thus (4a) is equivalently replaced by (4b) (notice that in (4b), $b_{C,j,\kappa} = 1$ implies $z_{C,j,\kappa} = v_j$, and $b_{C,j,\kappa} = 0$ implies $z_{C,j,\kappa} = 0$):

$$\begin{cases} q_{C,j} = \sum_\kappa CP_{j,\kappa} z_{C,j,\kappa} \\ v_j - (1 - b_{C,j,\kappa})\bar{v}_j \leq z_{C,j,\kappa} \leq v_j - (1 - b_{C,j,\kappa})\underline{v}_j \\ b_{C,j,\kappa} \underline{v}_j \leq z_{C,j,\kappa} \leq b_{C,j,\kappa} \bar{v}_j,\ \sum_\kappa b_{C,j,\kappa} \leq 1, b_{C,j,\kappa} \in \{0,1\} \end{cases} \quad (4b)$$

As for the OLTC in branch $(i,j)$, suppose $t_{ij}$ has $\mathrm{K}_{T,ij}$ possible values, denoted by $TR_{ij,\kappa}$ for $\kappa = 1, \ldots, \mathrm{K}_{T,ij}$. Similarly to (4a), with $\mathrm{K}_{T,ij}$ binary variables $b_{T,ij,\kappa}$, $t_{ij}^2 v_i$ is formulated as

$$\begin{cases} t_{ij}^2 v_i = \sum_\kappa TR_{ij,\kappa}^2 b_{T,ij,\kappa} v_i \\ \sum_\kappa b_{T,ij,\kappa} = 1,\ b_{T,ij,\kappa} \in \{0,1\} \end{cases}. \quad (5)$$

Similarly to the transformation in (4b), if both $v_i$ and $b_{T,ij,\kappa}$ are optimal variables, then each bilinear term $b_{T,ij,\kappa} v_i$ should be replaced by an auxiliary variable $z_{T,ij,\kappa}$ subject to the McCormick's constraints. Notice, however, that for the OLTCs in the distribution substation, since $v_1 = v^{\text{set}}$ is an (uncertain) input, the constraints (5) are linear in $b_{T,1j,\kappa}$.

Finally, for the RPP problem, there is still an additional constraint: the boundary-bus reactive power is specified by the TSO, say $q_B$, so as to balance the TPS-side reactive power. Thus, the reactive power constraint for node 1 in (1b) should be modified as

$$q_1 = q_B + q_{C,1} = \sum_{k \in \mathcal{T}(1)} q_{1k}. \quad (6)$$

Moreover, if $q_B$ is an arbitrary value in the range $[\underline{q}_B, \bar{q}_B]$, it follows that

$$q_B = \underline{q}_B + u_Q(\bar{q}_B - \underline{q}_B) \text{ where } u_Q \in [0,1]. \quad (7)$$

The constraints (1)-(7) describe a deterministic DPS's operation model, based on which a two-stage RO-based RPP estimation problem will be established.

## III. Two-Stage RO Model

### A. Variable Partition

As explained in Section I, for this RO-based RPP estimation, the uncertainty variables $u$ include the uncertain DERs' active power $p_{G,j}$, the boundary-bus voltage $v^{\text{set}}$ that is uncertain to the DSO when the latter computes the RPP, and the uncertain reactive power $q_B$ that will be required by the TSO and lies in the range $[\underline{q}_B, \bar{q}_B]$. However, for an easier solution to the RO, for a fixed $[\underline{q}_B, \bar{q}_B]$, the uncertain $q_B$ should be replaced by an uncertain $u_Q \in [0,1]$, as per (7) [22]. Hence, in the sequel, a $u_Q \in [0,1]$ rather than $q_B$ is included in $u$.

In the first stage, the DSO should pre-estimate the RPP, denoted by the range $[\underline{q}_B, \bar{q}_B]$: here, $\underline{q}_B$ (or $\bar{q}_B$) > (resp. <) 0 means *inductive* (resp. *capacitive*), so the first-stage variables $x$ include $\underline{q}_B$, $\bar{q}_B$. Additionally, following the assumption in [17],[23],[24], the discrete optimal variables $b_{C,j,\kappa}$ and $b_{T,ij,\kappa}$ are also included in $x$, implying that they should be determined in the first stage and remain unchanged until the next update of the RPP (e.g., ten minutes later). This would inhibit repeatedly regulating the OLTCs and the shunts, which is usually a practical requirement in field operation.

The second stage is to check whether any reactive power in $[\underline{q}_B, \overline{q}_B]$ is realizable for every possible instance of $u$ under conditions of a fixed $x$. Hence, the second-stage variables $y$ include the continuous control variables $q_{G,j}$, the network variables $v_j$, $p_{ij}$, $q_{ij}$ and $l_{ij}$, and the auxiliary variables $z_{C,j,\kappa}$, $z_{T,ij,\kappa}$. Mathematically, if the answer to the STP is yes, then there should exist a $y$ associated with an instance of $u$ under conditions of the fixed $x$.

In short, in the RO model below, we have three types of variables: $u = \left[[p_{G,j}]; v^{\text{set}}; u_Q\right]$, $x = \left[[b_{C,j,\kappa}]; [b_{T,ij,\kappa}]; \underline{q}_B; \overline{q}_B\right]$, and $y = \left[[q_{G,j}]; [v_j]; [p_{ij}]; [q_{ij}]; [l_{ij}]; [z_{C,j,\kappa}]; [z_{T,ij,\kappa}]\right]$, where $[\,]$ denotes a vector of the corresponding variables.

### B. Robust Formulation

The minimization in (8) is to maximize the RPP $[\underline{q}_B, \overline{q}_B]$:

$$\text{Min} \quad f(x) = \left(\underline{q}_B - \underline{Q}\right)^2 + \left(\overline{q}_B - \overline{Q}\right)^2 \quad (8)$$

$$\text{where} \quad \underline{Q} \leq \underline{q}_B \leq \overline{q}_B \leq \overline{Q}. \quad (9)$$

Here, $\underline{Q}$ and $\overline{Q}$ are the predefined limits for the boundary-bus reactive power, e.g., the capacity of the transformer in the distribution substation.

Following the definitions of $u$, $x$ and $y$, (1)-(7) and (9), where the bilinear terms are eliminated via (4b) and (5), can be written in a condensed form:

$$\begin{cases} \left\|\mathbf{A}_l u + \mathbf{B}_l y\right\|_2 = \mathbf{f}_l^T y + \mathbf{h}_l^T u, \forall l \in \mathcal{E} \\ \mathbf{E}u + \mathbf{F}y \leq \mathbf{g}, \quad \mathbf{G}u + \mathbf{H}y = \mathbf{d} \\ \mathbf{C}x + \mathbf{D}y \leq \mathbf{b}, \quad \mathbf{J}x \leq \mathbf{s} \end{cases}, \quad (10)$$

where the bold capital (resp. small) letters denote the coefficient matrices (resp. coefficient vectors) in the constraints.

Moreover, we assume that the uncertainty is constrained by a polyhedron, i.e., $u \in \mathcal{U} = \{u | \mathbf{R}u \leq \mathbf{r}\}$. A simple construction of this polyhedron is, but is not limited to, the following:

$$\mathcal{U} = \left\{ u \left| \begin{array}{l} p_{G0,j} - \alpha_{G,j}\Delta_{G,j} \leq p_{G,j} \leq p_{G0,j} + \alpha_{G,j}\Delta_{G,j} \\ \underline{v}^{\text{set}} \leq v^{\text{set}} \leq \overline{v}^{\text{set}}, \quad 0 \leq u_Q \leq 1 \end{array} \right. \right\}, \quad (11)$$

where $p_{G0,j}$ denotes the active power of the DER at node $j$ at the moment when the DSO computes the RPP; $\Delta_{G,j}$ denotes the maximal possible deviation of the DER's active power in the next several (e.g., ten) minutes; $\alpha_{G,j}$ is used to regulate the conservativity of the uncertainty set, and a smaller value means less uncertainty in $p_{G,j}$; and $\underline{v}^{\text{set}}$ and $\overline{v}^{\text{set}}$ represent the range in which the $v^{\text{set}}$ can be located, which can be estimated by the DSO via historical data.

Then, the model of this robust RPP evaluation problem is formulated below:

$$\min_x f(x)$$
$$\text{s.t.} \; \mathbf{J}x \leq \mathbf{s},$$
$$\forall u \in \mathcal{U}, \exists y: \begin{cases} \left\|\mathbf{A}_l u + \mathbf{B}_l y\right\|_2 = \mathbf{f}_l^T y + \mathbf{h}_l^T u, \forall l \in \mathcal{E} \\ \mathbf{E}u + \mathbf{F}y \leq \mathbf{g}, \; \mathbf{G}u + \mathbf{H}y = \mathbf{d}, \\ \mathbf{C}x + \mathbf{D}y \leq \mathbf{b} \end{cases} \quad (12)$$

The problem (12) is a two-stage RO problem [25], which indicates that for a given $x$, one needs to check whether there is a feasible $y$ for an arbitrary instance of $u \in \mathcal{U}$ (defined as the SSP). If the answer is yes, we call this $x$ *"robust feasible"*; otherwise, we need to search for a new $x$ (defined as the FSP).

### C. Transformation for Tractability

*1) Conic Relaxation*

The SSP in (12) is difficult to solve because of the nonconvexity in the constraints. For tractability, a conic-relaxation-based maximin problem is formulated instead, as shown below:

$$\mathcal{Q}(x) = \max_{u \in \mathcal{U}} \min_y \sum_{(i,j)} \rho_{ij} l_{ij}$$
$$\text{s.t.} \begin{cases} \left\|\mathbf{A}_l u + \mathbf{B}_l y\right\|_2 \leq \mathbf{f}_l^T y + \mathbf{h}_l^T u, \forall l \in \mathcal{E} \\ \mathbf{E}u + \mathbf{F}y \leq \mathbf{g}, \; \mathbf{G}u + \mathbf{H}y = \mathbf{d}, \; \mathbf{D}y \leq \mathbf{b} - \mathbf{C}x \end{cases} \quad (13)$$

where $\rho_{ij}$ are positive numbers, like $1$. Let $\tilde{c}^T y$ denote the objective function in (13). Two comments are provided below:

First, the constraints in (13) are the conic relaxation of those in the SSP. This implies that for an instance of $u$, any $y$ infeasible for (13) must be infeasible for the SSP.

Secondly, based on the proof in [26],[27], under some sufficient conditions (e.g., the objective in (13) is increasing in $l_{ij}$ )[2], for an instance of $u$, the equality in the SOC constraints in (13) holds for an optimal $y$. Thus, this optimal $y$, for which $\tilde{c}^T y$ is finite, is feasible for the constraints in the SSP under the same $x$ and the same instance of $u$.

*2) Dualization*

Problem (13) is further transformed into a max-max problem via dualization, as follows:

$$\max_{u \in \mathcal{U}} \max_{\xi, \omega, \vartheta, \eta, \lambda_{l \in \mathcal{E}}, \pi_{l \in \mathcal{E}}} \xi^T u + (\mathbf{C}x - \mathbf{b})^T \vartheta - \mathbf{g}^T \omega - \mathbf{d}^T \eta$$
$$\text{s.t.}(\mathcal{W}) \begin{cases} \mathbf{G}^T \eta + \mathbf{E}^T \omega + \sum_{l \in \mathcal{E}} \left(\mathbf{A}_l^T \pi_l - \mathbf{h}_l \lambda_l\right) - \xi = 0 \\ \mathbf{H}^T \eta + \mathbf{F}^T \omega + \mathbf{D}^T \vartheta + \sum_{l \in \mathcal{E}} \left(\mathbf{B}_l^T \pi_l - \mathbf{f}_l \lambda_l\right) = -\tilde{c} \\ \omega \geq 0, \; \vartheta \geq 0, \; \left\|\pi_l\right\|_2 \leq \lambda_l, \; \forall l \in \mathcal{E} \end{cases} \quad (14)$$

where $\eta$, $\omega$ and $\vartheta$ are the multipliers associated with the linear equalities and inequalities in (13), respectively; the pairs $(\pi_l, \lambda_l)$ are the multipliers associated with the SOC inequalities in (13); and $\xi$ is an auxiliary variable to simplify the bilinear terms brought in by dualization.

Although (14) is a bilinear optimization problem, the sets

---

[2] Empirically and also as verified in the studies [17] and [21], even when the conditions in [26],[27] are not satisfied, the equality in the SOC constraints still often holds when a large $\rho_{ij}$ is used.

$\mathcal{W}$ and $\mathcal{U}$ are distinct and the sequence of the outer- and inner-level problems is exchangeable. By exchanging the sequence [28], (14) is then transformed into $\max_{(\xi,\omega,\vartheta,\eta,\lambda_{l\in\mathcal{E}},\pi_{l\in\mathcal{E}})\in\mathcal{W}}(\mathbf{C}x-\mathbf{b})^{\mathrm{T}}\vartheta - \mathbf{g}^{\mathrm{T}}\omega - \mathbf{d}^{\mathrm{T}}\eta + \max_{u\in\mathcal{U}} \xi^{\mathrm{T}}u$, where the inner-level problem is a linear programming over $u$ with a given $\xi$. Since $\mathcal{U}$ is a nonempty polyhedron, which implies that the Slater condition is satisfied, it follows that maximal $\xi^{\mathrm{T}}u = \mathbf{r}^{\mathrm{T}}\phi$, where $\phi$ is subject to

$$\mathbf{R}u \leq \mathbf{r}, \phi \geq 0, \phi^{\mathrm{T}}(\mathbf{r}-\mathbf{R}u)=0, \xi = \mathbf{R}^{\mathrm{T}}\phi, \quad (15a)$$

or equivalently, a system of SOS1 variables:

$$\mathbf{R}u \leq \mathbf{r}, \phi \geq 0,, \xi = \mathbf{R}^{\mathrm{T}}\phi, \phi_i = b_{1,i}, (\mathbf{r}-\mathbf{R}u)_i = b_{2,i}, \quad (15b)$$

where subscript $i$ denotes the $i$-th element of the vector, and SOS1 means that for every pair $(b_{1,i}, b_{2,i})$, at most one of them is nonzero (cf. [29]).

Thus, Problem (13) is transformed into a mixed-integer SOC programming below:

$$\mathcal{Q}(x) = \max_{b_1,b_2,u,\phi,\omega,\vartheta,\eta,\lambda_{l\in\mathcal{E}},\pi_{l\in\mathcal{E}}} (\mathbf{C}x-\mathbf{b})^{\mathrm{T}}\vartheta - \mathbf{g}^{\mathrm{T}}\omega - \mathbf{d}^{\mathrm{T}}\eta + \mathbf{r}^{\mathrm{T}}\phi$$

$$\text{s.t.} \begin{cases} \mathbf{G}^{\mathrm{T}}\eta + \mathbf{E}^{\mathrm{T}}\omega + \sum_{l\in\mathcal{E}}\left(\mathbf{A}_l^{\mathrm{T}}\pi_l - \mathbf{h}_l\lambda_l\right) - \mathbf{R}^{\mathrm{T}}\phi = 0 \\ \mathbf{H}^{\mathrm{T}}\eta + \mathbf{F}^{\mathrm{T}}\omega + \mathbf{D}^{\mathrm{T}}\vartheta + \sum_{l\in\mathcal{E}}\left(\mathbf{B}_l^{\mathrm{T}}\pi_l - \mathbf{f}_l\lambda_l\right) = -\tilde{c} \\ \mathbf{R}u \leq \mathbf{r}, \phi \geq 0,, \xi = \mathbf{R}^{\mathrm{T}}\phi, \phi_i = b_{1,i}, (\mathbf{r}-\mathbf{R}u)_i = b_{2,i} \\ \omega \geq 0, \ \vartheta \geq 0, \ \|\pi_l\|_2 \leq \lambda_l, \forall l \in \mathcal{E}, \ (b_{1,i}, b_{2,i}): \text{SOS1} \end{cases} \quad (16)$$

Problem (16) is easy to solve via off-the-shelf tools like CPLEX. Moreover, via the techniques shown in [29] and [30], (16) can be further transformed into a continuous optimization problem that typically can be efficiently solved.

The following comments indicate how (16) is associated with the SSP in (12), which is the basis of the solution strategy in Section IV:

- As per the weak duality theorem, $\mathcal{Q}(x) = +\infty$ implies that for an $x$, there is an instance of $u$ under which there is no feasible $y$. Thus, the answer to the associated SSP is no, and then a new $x$ should be found via the FSP [3].

- If (16) is infeasible, i.e., $\mathcal{Q}(x) = -\infty$, it can be inferred that the dual of the inner-level problem of (13) must be infeasible for any instance of $u$ (notice that $\mathcal{U}$ is nonempty). This is because otherwise the inner-level problem of (13) must be unbounded as per the duality theorem, which contradicts the fact that $\tilde{c}^{\mathrm{T}}y$ is bounded from both sides (cf. (1f) and (13)). Hence, the FSP should be invoked to update $x$.

- If $\mathcal{Q}(x)$ is finite, following the second comment under (13), the answer to the associated SSP must be yes, so this $x$ is robust feasible.

---

[3] Indeed, $\mathcal{Q}(x) > \sum_{(i,j)} \rho_{ij}|I_{ij}|^2$ implies that this condition occurs, because $\sum_{(i,j)} \rho_{ij}|I_{ij}|^2$ is an upper bound of $\tilde{c}^{\mathrm{T}}y$, which should have been larger than $\mathcal{Q}(x)$ if both (16) and the inner-level problem of (13) were feasible.

## IV. SOLUTION STRATEGY

The C&CG algorithm proposed in [19] is intended for solving a two-stage RO problem of the format, $\min_{x\in\mathcal{X}} f(x) + \max_{u\in\mathcal{U}} \min_{y\in\mathcal{F}(x,u)} g(y)$, where $\mathcal{X}$ denotes the feasible set regarding $x$, $g(y)$ denotes the objective of the SSP, and $\mathcal{F}(x,u)$ denotes the feasible set regarding $y$ with fixed $x, u$. The C&CG is an iterative algorithm: briefly, in every iteration, with a given $x$, a subproblem (SP) $\max_{u\in\mathcal{U}} \min_{y\in\mathcal{F}(x,u)} g(y)$ is solved to generate new *cuts* that will be added in a master problem (MP) either to improve the optimality of the solution or to cut off a part of $\mathcal{X}$ that leads to an infeasible SP; then, the MP is updated and solved for a new $x$. It is proved in [19] that if the uncertainty set $\mathcal{U}$ is a polyhedron, "the C&CG algorithm will converge to the optimal value of the two-stage RO in finite iterations."

As (12) is a two-stage RO problem, the C&CG algorithm is applicable. Nevertheless, owing to the nonconvexity in the SSP, (16) is solved in the C&CG algorithm, and the *cuts* are generated on the basis of the aforementioned relations between (16) and the SSP. The C&CG-based solution strategy is stated below:

---

**Algorithm 1**

**Step 1**: Set $UB = +\infty$, $LB = -\infty$, convergence error $\varepsilon$, penalty $\beta$, and $k = 0$.

**Step 2**: Solve the following MP:

$$\min_{x,\nu,y_r} \left(\underline{q}_{\mathrm{B}} - \underline{Q}\right)^2 + \left(\overline{q}_{\mathrm{B}} - \overline{Q}\right)^2 + \beta \cdot \nu \quad (17a)$$

$$\text{subject to} \quad \mathbf{J}x \leq \mathbf{s}, \quad (17b)$$

$$\nu \geq \tilde{c}^{\mathrm{T}}y_r, \quad \forall r \leq k \quad (17c)$$

$$\begin{cases} \|\mathbf{A}_l u_r^* + \mathbf{B}_l y_r\|_2 \leq \mathbf{f}_l^{\mathrm{T}}y_r + \mathbf{h}_l^{\mathrm{T}}u_r^*, \forall l \in \mathcal{E} \\ \mathbf{E}u_r^* + \mathbf{F}y_r \leq \mathbf{g}, \ \mathbf{G}u_r^* + \mathbf{H}y_r = \mathbf{d}, \ \mathbf{C}x + \mathbf{D}y_r \leq \mathbf{b} \end{cases} \forall r \leq k \quad (17d)$$

Derive an optimal solution $x_{k+1}^*$ and denote the optimum by $f_{\mathrm{MP},k+1}$; update $LB = f_{\mathrm{MP},k+1}$.

**Step 3**: Solve the SP (16). If it is feasible and $\mathcal{Q}(x_{k+1}^*) \leq \sum_{(i,j)} \rho_{ij}|I_{ij}|^2$, update $UB = \min\{UB, f(x_{k+1}^*) + \beta \cdot \mathcal{Q}(x_{k+1}^*)\}$.

**Step 4**: If $UB - LB \leq \varepsilon$, return $[\underline{q}_{\mathrm{B},k+1}^*, \overline{q}_{\mathrm{B},k+1}^*]$; otherwise, do the following:

(a) If $\mathcal{Q}(x_{k+1}^*) \leq \sum_{(i,j)} \rho_{ij}|I_{ij}|^2$, record this $u_{k+1}^*$, and create new optimal variables $y_{k+1}$ in the MP and add the following constraints:

$$\nu \geq \tilde{c}^{\mathrm{T}}y_{k+1} \quad (18a)$$

$$\begin{cases} \|\mathbf{A}_l u_{k+1}^* + \mathbf{B}_l y_{k+1}\|_2 \leq \mathbf{f}_l^{\mathrm{T}}y_{k+1} + \mathbf{h}_l^{\mathrm{T}}u_{k+1}^*, \forall l \in \mathcal{E} \\ \mathbf{E}u_{k+1}^* + \mathbf{F}y_{k+1} \leq \mathbf{g}, \ \mathbf{G}u_{k+1}^* + \mathbf{H}y_{k+1} = \mathbf{d}, \ \mathbf{C}x + \mathbf{D}y_{k+1} \leq \mathbf{b} \end{cases} \quad (18b)$$

(b) If $\mathcal{Q}(x_{k+1}^*) > \sum_{(i,j)} \rho_{ij}|I_{ij}|^2$, record this $u_{k+1}^*$, and create new optimal variables $y_{k+1}$ in the MP and add a constraint similar to (18b).

(c) If (16) is infeasible, then select from $\mathcal{U}$ a $u_{k+1}^*$ that is different from $u_1^*, \ldots, u_k^*$. Record this $u_{k+1}^*$, and create new optimal variables $y_{k+1}$ in the MP and add there a constraint similar to (18b).

**Step 5**: Update $k = k + 1$ and go to Step 2.

Notice that Algorithm 1 converges to an optimum $f(x)+\beta\tilde{c}^\mathrm{T}y$ in finite iterations. Relative to (12), the additional term $\beta\tilde{c}^\mathrm{T}y$ is there to ensure that the equality in the SOC constraints holds at the optimizer of the MP. Certainly, the second term should be much smaller than the first one, so one can first try a small $\beta$, e.g., $10^{-3}$, and then increase it until the equality in the SOC constraints holds.

In practice, Algorithm 1 can be terminated the first time $\mathcal{Q}(x_{k+1}^*) \leq \sum_{(i,j)} \rho_{ij} |I_{ij}|^2$ appears. This is because with a small $\beta$, $f_{\mathrm{MP},k} \approx f(x_k^*)$ and then the range $[\underline{q}_{\mathrm{B},k}^*, \overline{q}_{\mathrm{B},k}^*]$ shrinks with an increasing $k$, as more cuts are added in the MP. Hence, when the algorithm goes to Step 4a for the first time, it means that the current $[\underline{q}_{\mathrm{B},k}^*, \overline{q}_{\mathrm{B},k}^*]$ is already robust feasible[4] and thus there is no need to shrink this range any longer to achieve the optimum $f(x)+\beta\tilde{c}^\mathrm{T}y$, which is more or less influenced by the second term.

Finally, if the DistFlow equations in (12) are replaced by linearized single- or three-phase power flow equations, (12) becomes a two-stage linear RO problem[5], but Algorithm 1 is still workable for such a problem as per the proof in [19].

## V. Case Studies

### A. Simulation Systems

The first test system is a node-5 DPS whose data are available in [32]. There is a 1.4-MVA DER and a continuously regulatable 1-MVar SVG/SVC at node 3. The active power of the DER varies in the range [0.4, 0.6] MW. As per the data in [32], there are two switchable capacitor bank shunts at nodes 3 and 4, and we assume that either of them has three switchable banks whose individual capacity is 0.4 p.u., namely, 0.4 MVar at a 1.0-pu voltage. In addition, there is an OLTC in the distribution substation and the ratio is regulatable in the range [0.98, 1.02] with a step 0.01.

The second test system is the well-known node-33 system. There are five 1.1-MVA DERs at nodes 3, 5, 11, 20 and 25; their active power varies in the range [0.32, 0.48] MW. There are four switchable capacitor shunts at nodes 7, 19, 27 and 33; each of them has three banks whose individual capacity is 0.2 p.u. The tap of the OLTC in the distribution substation is regulatable and the possible ratio is in the range [0.98, 1.06] with a step 0.01.

The third test system is a node-77 system whose data are available in [33], but the nodes are numbered from 1 to 77 here. There are eight 2.1-MVA DERs at nodes 4, 7, 10, 14, 18, 20, 21 and 23, and the active power of each DER varies in the range [0.55, 0.85] MW. There are two continuously regulatable 1-MVar SVG/SVCs at nodes 29 and 31 and four switchable capacitor shunts at nodes 6, 18, 39 and 77. Every capacitor shunt has three banks whose individual capacity is 0.3 p.u. The tap of the OLTC in the distribution substation is regulatable and the possible ratio is in the range [0.98, 1.06] with a step 0.01.

For all three test systems, the boundary-bus voltage setpoint is assumed to vary in the range [0.99, 1.01] p.u. MATLAB and CPLEX are used as the simulation environment and the solver, respectively.

### B. Benchmark Method for Comparison

To show the benefit of this robust RPP estimation method, we compare it with a commonly used DM[6], the idea of which is shown in [3]-[5]. Briefly, in the DM, the RPP $[\check{\underline{q}}_\mathrm{B}, \widehat{\overline{q}}_\mathrm{B}]$ is solved via two OPF problems: first, let $p_{G,j} = p_{G0,j}$, $v^{\mathrm{set}} = 1$ and $q_\mathrm{B} = \underline{q}_\mathrm{B}$ (record this as $\check{u}$), then solve an optimal $\check{\underline{q}}_\mathrm{B}$ via (19a); second, let $p_{G,j} = p_{G0,j}$, $v^{\mathrm{set}} = 1$ and $q_\mathrm{B} = \overline{q}_\mathrm{B}$ (record this as $\widehat{u}$), then solve an optimal $\widehat{\overline{q}}_\mathrm{B}$ via (19b).

$$\min_{x,y} \left(\underline{q}_\mathrm{B} - Q\right)^2$$
$$\text{s.t.} \begin{cases} \mathbf{J}x \leq \mathbf{s}, \ \|\mathbf{A}_l\check{u} + \mathbf{B}_l y\|_2 = \mathbf{f}_l^\mathrm{T} y + \mathbf{h}_l^\mathrm{T}\check{u}, \forall l \in \mathcal{E} \\ \mathbf{E}\check{u} + \mathbf{F}y \leq \mathbf{g}, \ \mathbf{G}\check{u} + \mathbf{H}y = \mathbf{d}, \ \mathbf{C}x + \mathbf{D}y \leq \mathbf{b} \end{cases} \quad (19\mathrm{a})$$

$$\min_{x,y} \left(\overline{q}_\mathrm{B} - Q\right)^2$$
$$\text{s.t.} \begin{cases} \mathbf{J}x \leq \mathbf{s}, \ \|\mathbf{A}_l\widehat{u} + \mathbf{B}_l y\|_2 = \mathbf{f}_l^\mathrm{T} y + \mathbf{h}_l^\mathrm{T}\widehat{u}, \forall l \in \mathcal{E} \\ \mathbf{E}\widehat{u} + \mathbf{F}y \leq \mathbf{g}, \ \mathbf{G}\widehat{u} + \mathbf{H}y = \mathbf{d}, \ \mathbf{C}x + \mathbf{D}y \leq \mathbf{b} \end{cases} \quad (19\mathrm{b})$$

### C. Results from the Robust and Deterministic Methods

The RPPs of the robust method and the DM are compared in Table I and the optimal decisions on the status of the capacitor shunts and the OLTC's tap ratios are listed in Tables II and III, respectively.

TABLE I
RPP COMPARISON (UNIT: MVAR)

| System | Robust Method | DM |
|---|---|---|
| Node-5 | [0.39, 5.17] | [0.33, 7.69] |
| Node-33 | [-4.34, 4.26] | [-5.66, 6.26] |
| Node-77 | [-3.69, 6.65] | [-13.08, 20.62] |

TABLE II
VALUES OF CAPACITOR SHUNTS

| System | Robust Method | DM via (19a) | DM via (19b) |
|---|---|---|---|
| Node-5 | [1.2, 1.2] | [1.2, 1.2] | [0, 0] |
| Node-33 | [0.6, 0.4, 0.6, 0.6] | [0.6, 0.6, 0.6, 0.6] | [0, 0, 0, 0.6] |
| Node-77 | [0.9, 0.9, 0.9, 0.9] | [0.9, 0.9, 0.9, 0.9] | [0, 0, 0, 0.3] |

TABLE III
OLTC TAP RATIOS

| System | Robust Method | DM via (19a) | DM via (19b) |
|---|---|---|---|
| Node-5 | 1.02 | 1.02 | 1.02 |
| Node-33 | 1.03 | 1.03 | 1.05 |
| Node-77 | 1.03 | 1.00 | 1.06 |

Table I indicates that the RPP yielded by the robust method is generally a subset of that yielded by the DM. Hence, the DM may produce an RPP that is not completely reliable. This assertion can be substantiated by the following example: for the node-5 system, let $[\underline{q}_B, \overline{q}_B] = [0.33, 0.39]$ MVar and the values of the capacitor shunts and tap position be [1.2, 1.2] p.u. and 1.02 respectively; then solve (16). It is found that $\mathcal{Q}(x) > \sum_{(i,j)} \rho_{ij} |I_{ij}|^2$, which implies that the SSP is infeasible under this $[\underline{q}_B, \overline{q}_B]$. This indicates that even if the DM and the robust method yield the same decision about the status of the discrete control variables, the RPP estimated via the DM is likely to include the reactive power that cannot be actually realized via the control in the DPS.

---

[4] Technically, this robust feasibility is based on the premise that the equality in the SOC constraints in (13) holds for the optimal $y$. This can usually be achieved by setting large $\rho_{ij}$, as explained in Footnote 2.

[5] It is still an open question whether a linearized power flow model or an SOC relaxed DistFlow model is generally more accurate [31].

[6] Although the scenario-sample-based stochastic programming is also popularly used to solve an uncertain problem, we think it is inapt for the RPP estimation because only limited instances of $u$ rather than all possible instances can be considered there, which makes the result not completely robust feasible. Hence, we will only compare the robust method and the DM.

We think there are two reasons for the difference in Table I. The first, as shown in Tables II and III, is that in the DM, $\breve{q}_B$ and $\hat{q}_B$ may be achieved at different optimum discrete variables. This will exaggerate the actual RPP of a DPS, if the shunts and the tap positions cannot be punctually regulated in response to every TSO's reactive power requirement, which is often the case in field operation. The second reason is that the uncertainty is not considered in the DM. As in the above example, the range [0.33, 0.39] is not robust infeasible when the uncertainty in the DERs' active power and boundary-bus voltage is considered.

Therefore, relative to the DM, it is advantageous to adopt the proposed robust method for a completely reliable estimate of the RPP. In addition, as shown in Table IV, the robust method yields the reliable RPP in less than 5 iterations for all three systems, owing to the property that the C&CG algorithm converges in finite iterations for a polyhedron uncertainty set.

Moreover, the relative gaps regarding the conic relaxation, defined as $(b - \|a\|_2)/b$ for an SOC constraint $\|a\|_2 \leq b$, are listed in Table IV. The small numbers imply that the equality in the SOC constraints numerically nearly holds at the optimal solution, which accords with the observations in other DistFlow-based RO studies, e.g., [17].

TABLE IV
ITERATIONS AND GAPS IN THE ROBUST METHOD

| System | Node-5 | Node-33 | Node-77 |
|---|---|---|---|
| Iterations | 3 | 4 | 4 |
| Gaps | <10⁻¹⁰ | 4.7×10⁻⁵ | <10⁻¹⁰ |

### D. Factors Impacting RPP

It is often beneficial for a DSO to know what factors may affect the RPP. Although it is easy to understand that the length of the RPP would be greater if the capacity of the SVG/SVCs increased, it is not so obvious how the capacity of the capacitor shunts and the range of the tap ratio affect the RPP. It is also interesting to know how the uncertainty in DERs impacts the RPP. This section will investigate the impact of these factors.

*1) Impact of Shunt's Capacity and Range of Tap Ratios*

First, in the node-5 system, let the individual capacity of every capacitor shunt be 0.3 and 0.5 MVar sequentially. Then the simulation is redone, and the results are listed in Table V. There is a general trend: if the on/off status of the capacitors is the same, then the larger the capacity of the capacitor, the smaller are $\overline{q}_B$ and $\underline{q}_B$. This means that the DPS appears more "capacitive" to the TPS, which can be understood from the power flow balance of the whole DPS, as the capacitor shunts can be considered as a capacitive reactive power source.

TABLE V
IMPACT OF CAPACITY OF CAPACITOR SHUNTS IN NODE-5 SYSTEM

| Capacity of a Bank | 0.3 | 0.4 | 0.5 |
|---|---|---|---|
| $[\underline{q}_B, \overline{q}_B]$ (MVar) | [1.00, 5.80] | [0.39, 5.17] | [-0.22, 4.54] |
| Capacitors (p.u.) | [0.9, 0.9] | [1.2, 1.2] | [1.5, 1.5] |
| Tap Ratios | 1.02 | 1.02 | 1.02 |

Next, we will test the impact of the range of the tap ratio of the OLTC in the distribution substation. For the node-5 system, we assume that the lowest tap ratio is 0.98 and the highest one is 1.02, 1.03, 1.04 or 1.05, and the step of the ratio is 0.01. The test results are shown in Table VI. The results show that as the upper bound on the tap ratio increases from 1.02 to 1.04, both $\overline{q}_B$ and $\underline{q}_B$ decrease, making the DPS's RPP appear more "capacitive." We think this is because a high tap ratio will increase the nodal voltages (cf. Fig. 1) and also the capacitive reactive power from the capacitor shunts (cf. (4a)).

TABLE VI
IMPACT OF TAP RANGE IN NODE-5 SYSTEM

| Tap Range | [0.98, 1.02] | [0.98, 1.03] | [0.98, 1.04] | [0.98, 1.05] |
|---|---|---|---|---|
| $[\underline{q}_B, \overline{q}_B]$ (MVar) | [0.39, 5.17] | [0.34, 5.12] | [0.29, 5.06] | [0.29, 5.06] |
| Tap Ratios | 1.02 | 1.03 | 1.04 | 1.04 |

*2) Impact of Uncertainty in DERs*

To investigate the impact of the uncertainty in the DERs' active power, let the pair $(p_{G0,j}, \Delta_{G,j})$ be (0.4, 0.4) MW for the node-33 system and (0.7, 0.75) MW for the node-77 system. The results are listed in Tables VII and VIII. It can be seen that the length of the RPP shrinks with a larger uncertainty in the DER's active power, because the latter causes a smaller regulatable range of the DER's reactive power as per the constraint (2) and also reduces the RPP of the DPS. However, as also revealed in these two tables, despite the uncertainty in the DERs' active power, a DPS can still provide a nontrivial RPP to balance the reactive power in the TPS, which demonstrates the DPS's efficacy as a local reactive power prosumer.

TABLE VII
IMPACT OF DERS' UNCERTAINTY IN NODE-33 SYSTEM

| $\alpha_{G,j}$ | 0.1 | 0.3 | 0.6 | 0.9 |
|---|---|---|---|---|
| $[\underline{q}_B, \overline{q}_B]$ (MVar) | [-5.38, 4.58] | [-3.48, 4.18] | [-2.57, 4.14] | [-2.16, 2.64] |
| Capacitors (p.u.) | [0.6, 0.6, 0.6, 0.6] | [0.4, 0, 0.6, 0.6] | [0.2, 0, 0.4, 0.6] | [0.4, 0.4, 0.4, 0.6] |
| Tap Ratios | 1.03 | 1.03 | 1.04 | 1.03 |

TABLE VIII
IMPACT OF DERS' UNCERTAINTY IN NODE-77 SYSTEM

| $\alpha_{G,j}$ | 0.2 | 0.3 | 0.5 |
|---|---|---|---|
| $[\underline{q}_B, \overline{q}_B]$ (MVar) | [-3.69, 6.65] | [-3.41, 6.29] | [-2.83, 5.79] |
| Capacitors (p.u.) | [0.9, 0.9, 0.9, 0.9] | [0.6, 0, 0.9, 0.9] | [0.6, 0.9, 0.9, 0.9] |
| Tap Ratios | 1.03 | 1.03 | 1.03 |

## VI. CONCLUSION

This paper proposes a two-stage RO-based RPP estimation method, where the uncertainty in the DERs and the boundary-bus voltage is considered. The RPP is computed in the first stage and its robust feasibility for any possible instance of the uncertainty is checked in the second stage. For tractability in the second stage, an alternative problem is designed, via conic relaxation and dualization, and solved. The relation between the solution to this alternative problem and the robust feasibility of the RPP is analyzed. Then, a C&CG algorithm is adopted as an iterative solution that solves this two-stage robust model in finite iterations. The comparison between this robust method and a DM demonstrates that this robust method is preferable, as it yields a completely reliable RPP, which means that any reactive power in this RPP is realizable. Moreover, it is further confirmed that even in the presence of uncertain DERs and boundary-bus voltage, the DPS is still a competent local reactive power prosumer to help maintain the reactive power balance in the TPS.

Future studies may include testing the effect of this robust method for an RPP estimation problem with linearized three-phase power flow equations, and applying linearization techniques to the SOC constraints in (16) for a faster solution.